\documentclass[journal]{IEEEtran}
\usepackage{subcaption}
\usepackage{cite}
\usepackage{amsmath,amssymb,amsfonts}
\usepackage{algorithmic}
\usepackage{graphicx}
\usepackage{textcomp}
\usepackage[table]{xcolor}
\usepackage{multirow}
\usepackage{multicol}
\usepackage{dblfloatfix}

\def\BibTeX{{\rm B\kern-.05em{\sc i\kern-.025em b}\kern-.08em
		T\kern-.1667em\lower.7ex\hbox{E}\kern-.125emX}}

\makeatletter

\makeatletter
\newcommand*\titleheader[1]{\gdef\@titleheader{#1}}
\AtBeginDocument{%
	\let\st@red@title\@title%
	\def\@title{%
		\bgroup\normalfont\large\centering\@titleheader\par\egroup
		\vskip0.1em\st@red@title}
}
\makeatother

\title{Inter-Slice Mobility Management in 5G: Motivations, Standard Principles, 
Challenges and Research Directions}
\titleheader{\scriptsize \textbf{\copyright 2022 IEEE. Personal use of this 
material is 
	permitted. 
	Permission from IEEE 
	must be obtained for all other uses, in any current or future media, 
	including reprinting/republishing this material for advertising or 
	promotional purposes, creating new collective works, for resale or 
	redistribution to servers or lists, or reuse of any copyrighted component 
	of this work in other works.}}
\begin{document}

	\author{Muhammad~Mohtasim~Sajjad,~\IEEEmembership{Member,~IEEE,}
	        Carlos~J.~Bernardos,
	        Dhammika~Jayalath,~\IEEEmembership{Senior~Member,~IEEE,}
	        and~Yu-Chu~Tian,~\IEEEmembership{Senior~Member,~IEEE} \vspace{-3em}
	        \thanks{\textit{Muhammad Mohtasim Sajjad, Dhammika Jayalath, and 
	Yu-Chu 
	        Tian are with Queensland University of Technology, Brisbane, 
	Australia. 
	        Carlos J. Bernardos is with Universidad Carlos III de Madrid, 
	        Madrid, 
	        Spain.}}}

	\maketitle
	\pagestyle{empty}
\IEEEtitleabstractindextext{%
	
\begin{abstract}
	Mobility management in a sliced 5G network introduces new and complex 
	challenges. In a network-sliced environment, user mobility has to be 
	managed among not only different base stations or access technologies but 
	also different slices. Managing user mobility among slices, or 
	inter-slice mobility, motivates the need for new solutions. 
	This article, presented as a tutorial, focuses on the 
	problem of inter-slice mobility from the perspective of 3GPP standards for 
	5G. It provides a detailed overview of the relevant 3GPP standard 
principles. Accordingly, key technical gaps, challenges, and  
	corresponding research directions are identified towards achieving seamless 
	inter-slice mobility within the current 3GPP network slicing framework. 
	\end{abstract}


}

\maketitle
\IEEEdisplaynontitleabstractindextext

%
\IEEEpeerreviewmaketitle

\section*{Introduction}
\label{sec:introduction}
\IEEEPARstart{N}etwork slicing enables simultaneous provisioning of diverse 
service types over the same physical infrastructure. Four service types are 
defined for network slicing in the Third Generation 
Partnership Project (3GPP) Release 16 specifications  \cite{23501}. These
include Ultra-Reliable Low Latency Communications (URLLC), 
Vehicle-to-Everything (V2X), Massive IoT (MIoT), and the conventional enhanced 
Mobile Broadband (eMBB). For several use cases within each service 
type, the 
3GPP specifications support the offering of communication services via a single or 
different network slices as shown in Fig. \ref{fig: 
CommunicationServicesANDSlices}. 
\begin{figure}[htb!]
	\begin{center}
		\includegraphics[scale=0.325]{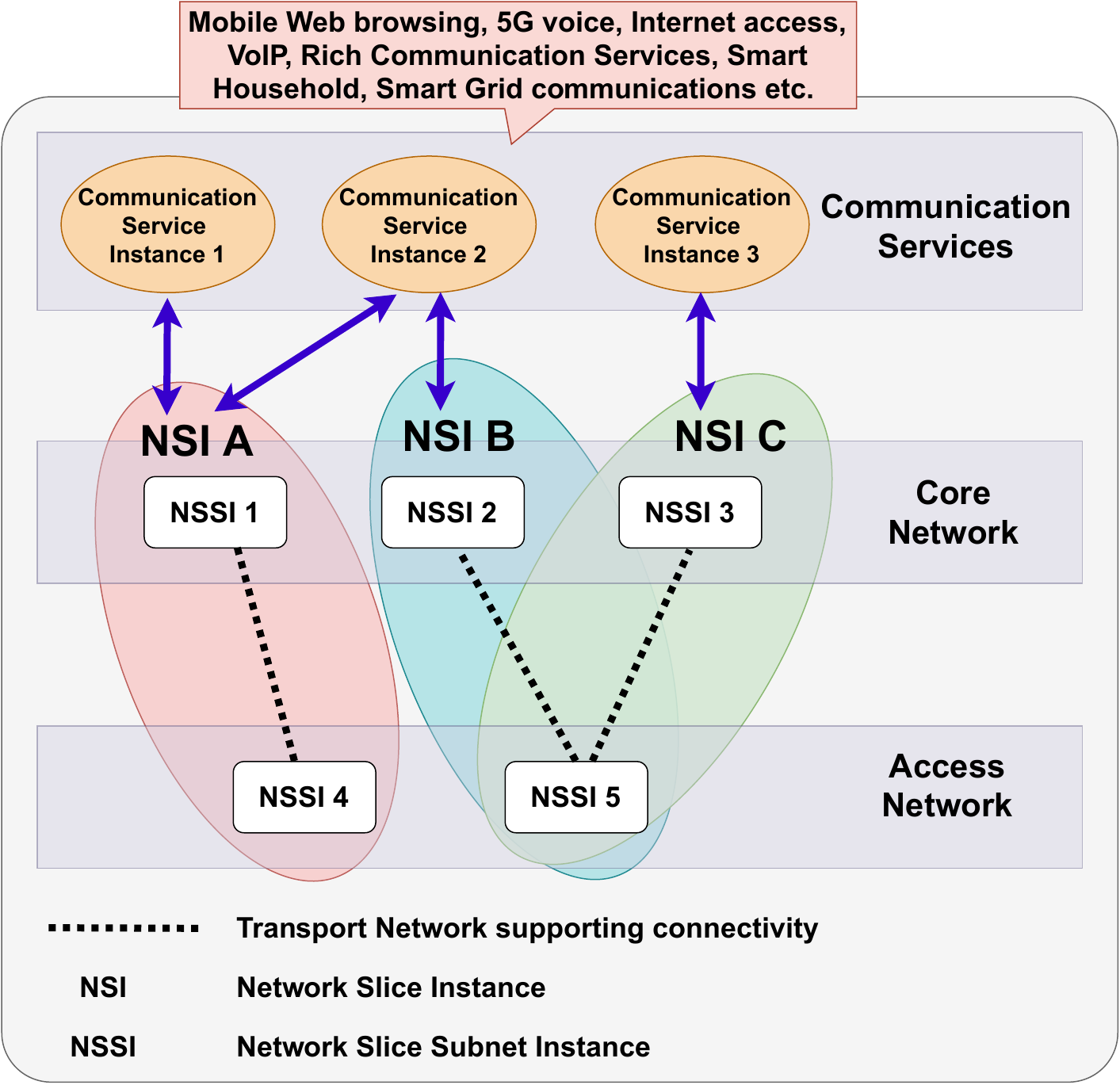}
		\caption{Communication services provided by different Network Slice 
		Instances (NSIs) as defined in the 3GPP standard TS 28.530. The shown 
		NSIs can be 
		different instances of the same or different slices. This paper 
		considers the latter and uses the generic term of network slice which 
		includes the definition of an NSI.}
		\label{fig: CommunicationServicesANDSlices}
	\end{center}
\end{figure}
\begin{figure}[h!]
	\begin{center}
		\includegraphics[scale=0.55]{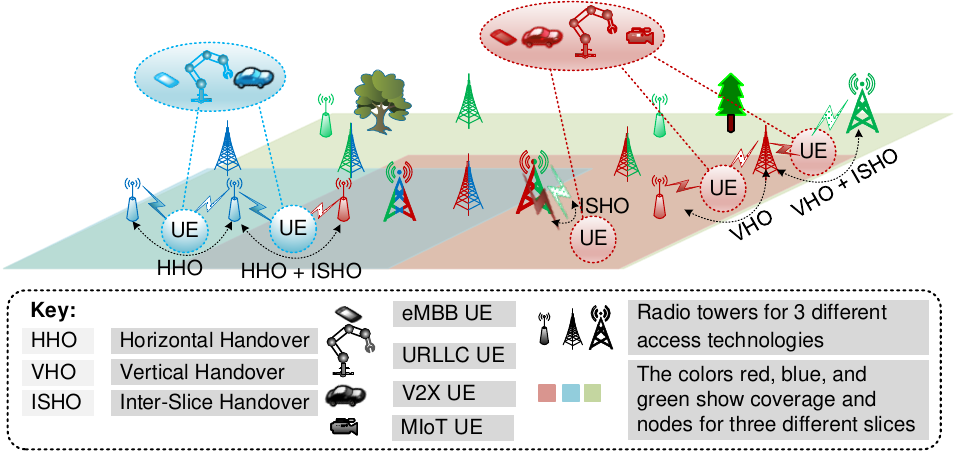}
		\caption{An example network-sliced mobility environment 
			showing 
			different handover types of UEs. The shown slices (and the UEs) 
			belong to one of the four service types (i.e.,
			URLLC, V2X, eMBB, and MIoT). }
		\label{fig: Scenario}
	\end{center}
\end{figure}


The availability of a communication service over different slices gives users a 
choice to change their slice if desired. Although a network slice is generally 
expected to deliver the user/service 
requirements consistently all the time (especially for URLLC and V2X use 
cases), users with active sessions may wish to change their slices if their 
preferences or requirements 
change over time. The slice owners may also wish to move users out of a slice, 
thus causing users to seek alternate slices to resume connectivity. 
Hence, in 
addition to the traditional horizontal (i.e., inter-cell/base-station 
handovers) and vertical handovers (i.e., inter-technology handovers), handovers 
among different slices (i.e., inter-slice handovers) are also expected in a 
network-sliced environment. An example scenario showing these different forms 
of handover is depicted in Fig. \ref{fig: Scenario}.


\begin{table*}[ht!]
	\centering
	\caption{Example inter-slice handover causes}
	\label{tab: ISHOCauses}
	{\rowcolors{2}{gray!20}{lightgray!20}
		\begin{tabular}{|p{1.6cm}|p{2.3cm}|p{8.2cm}|p{2.0cm}|}
			\hline
						\textbf{Main causes} & 
						\textbf{Potential triggers} 
						& 
						\textbf{Description} & 
						\textbf{Initiation 
								point}\\
			\hline
			%
			\cellcolor{white}& 
			\multirow{3}{*}{RAN conditions} & Due to the stochastic behavior 
			of 
			a 
			wireless channel, users in a mobile environment might 
			experience the drop in the Received 
			Signal Strength, Bit Error Rate, Signal-to-Interference and Noise 
			Ratio. & \multirow{3}{*}{UE-initiated} \\
			\cellcolor{white}{Slice-specific 
				conditions}& \multirow{3}{*}{Slice delay} & Depending on slice 
			composition and resources, factors such as link capacity at 
			front/backhaul, scheduling at RAN, queueing and NF processing 
			delays at core network can lead to undesirable slice delays. 
			& \multirow{3}{*}{UE-initiated}\\
			\cellcolor{white} & \multirow{2}{*}{Reliability} & The error rate 
			of a slice can increase (e.g., due to physical node/link failures, 
			security 
			attacks) resulting in reduced reliability of a slice. &	
			\multirow{2}{*}{UE-initiated}\\\hline
			
			\vspace{0.01mm}\cellcolor{white}Service/Appl. requirements & 
			\multirow{3}{*}{QoS 
				requirements} & Deterioration of the desired QoS of an ongoing 
			Application/Service (in terms of throughput, error rate, jitter 
			etc.) is possible due to different network events.  & 
			\multirow{3}{*}{UE-initiated}\\
			\hline
			\cellcolor{white} \vspace{7.5mm} Slice 
			owner/ & 
			\multirow{4}{*}{Slice 
				load} & The high utilization of available slice resources 
			can overload a slice. The slice owner/network operator can thus 
			enforce inter-slice mobility for some users, for instance, for 
			better 
			resource management, or to simply  
			serve their premium user base better. 
			& \multirow{4}{*}{Network-triggered}\\
			\cellcolor{white} Network \linebreak operator's \linebreak 
			preferences& 
			\multirow{3}{*}{Subscription policies}  & A network slice may 
			provide services to a user under specific subscription policies. 
			Once a user consumes its allowed services, it may be forced out 
			from the slice.   & \multirow{3}{*}{Network-triggered}\\
			\cellcolor{white}& \multirow{2}{*}{Pricing/Billing}  & A network 
			slice may 
			discontinue 
			its 
			services to a user, if a user runs out of its available credit. & 
			\multirow{2}{*}{Network-triggered}\\
			\hline
			\cellcolor{white}\vspace{1.5mm}Intra-/Inter- \linebreak Technology
			& 
			\multirow{3}{*}{Horizontal 
				handover} & A mobile user moving into a new Registration Area 
				might 
			move out of the coverage of its current slice, and may consequently 
			require to undergo inter-slice handover.  & 
			\multirow{3}{*}{UE-initiated}\\
			\cellcolor{white}handovers& 
			\multirow{3}{*}{Vertical handover} & The user’s choice to switch to 
			another access technology might also require it to undergo 
			inter-slice handover if its desired access technology is not 
			supported by the current slice. 
			& \multirow{3}{*}{UE-initiated}\\
			\hline
			\cellcolor{white}& Monetary costs & Different slices might offer 
			same 
			services at different costs.  & UE-initiated\\
			\cellcolor{white} & \multirow{3}{*}{Slice isolation level}  & Some 
			users might prefer slices with higher degree of isolation 
			characterized by the level of resource, infrastructure or NFs 
			sharing with other slices. & 
			\multirow{3}{*}{UE-initiated}\\
			\cellcolor{white}User preferences & \multirow{2}{*}{Slice security} 
			& Slices with strong 
			security mechanisms 
			might 
			be preferred by some users.  & \multirow{2}{*}{UE-initiated}\\
			\cellcolor{white} & \multirow{4}{*}{Slice policies} & 
			Slice owners would employ their own specific policies, which govern 
			their overall service and slice management. A user may prefer an 
			alternate slice if, for instance, frequenting between access 
			technologies, finds another slice offering suitable policies for 
			VHOs.  & \multirow{4}{*}{UE-initiated}\\
			\hline
			
		\end{tabular}
	}
\end{table*}

Inter-slice handover is a new form of handover. Unlike 
the horizontal handovers, the inter-slice handover may not always be 
event-triggered. Also, similar to vertical handovers, the inter-slice handovers 
may not always involve the physical 
mobility of the User Equipment (UE). Hence, the users/UEs belonging to any 
service type may require to undergo inter-slice handovers for a number of 
reasons. As shown in Fig. \ref{fig: Scenario}, the inter-slice handover may 
occur as a 
standalone event or as a result of a horizontal or a vertical handover. 

The communication services for different service types impose highly 
diverse requirements on the network. The tailored slices designed to meet these 
requirements will naturally have their own service-type specific inter-slice 
handover dynamics. Therefore, the inter-slice handover dynamics 
will be significantly divergent for slices belonging to different service 
types. 
For example, most of the MIoT UEs will be 
stationary or will have very low mobility \cite{NetSlicingSURVEY2017}, so the 
inter-slice handovers 
triggered due to horizontal handovers are less likely. Such scenarios, however, 
are expected to occur routinely for UEs belonging to, for instance, eMBB or V2X 
use cases. Likewise, the core URLLC and V2X slices are expected to be deployed 
closer to the UEs through edge technologies to achieve lower network 
delays. This may not always be the case for eMBB or MIoT UEs. Table \ref{tab: ISHOCauses} 
lists some example factors that can possibly trigger inter-slice mobility in 
different service types. In this article, however, we focus on the 
fundamental operational aspects of the inter-slice handovers without addressing 
the inter-slice mobility dynamics of any specific service type. 


It has been recognized that mobility management in a sliced network requires 
new protocols \cite{NetSlicingSURVEY2017, IntComputing2017}. However, in the 
contemporary research, 
only limited 
efforts have been made on the problem of inter-slice mobility management 
\cite{Zhang2017, Yousaf2017}. These solutions only give basic guidelines, and 
do not 
provide 
any specific framework or protocol for inter-slice mobility management.

The practical significance of inter-slice mobility management solutions 
requires them to comply with standard practices. In this regard, these solutions 
are required to be in compliance with the standard principles of the network 
slicing framework as specified by the 3GPP. The standard 3GPP network slicing 
framework, from architectural and operational perspective, is mainly 
concentrated on 5G core network, as will be discussed later. A novel 
Service-Based 
Architecture (SBA) at 5G core network provides the basis for network slicing. 
At the 
Radio Access Network (RAN) level, traffic/QoS differentiation among different 
service flows is applied to support slicing. 


\begin{figure*}[th!]
	\begin{center}
		\includegraphics[scale=0.15]{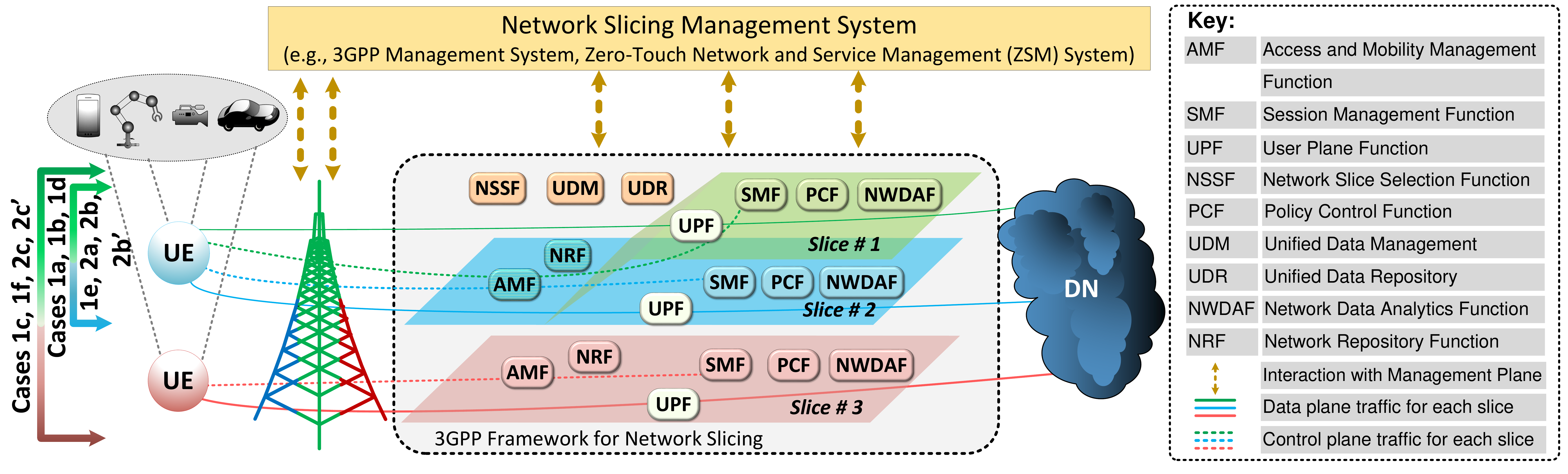}
		\caption{A representation of network slicing in 3GPP SBA (based on 
			\cite{3GPPNetSlicing}). Possible forms (or cases) of inter-slice 
			switching are 
			also shown, which are discussed in Section \ref{sec: forms}.}
		
		\label{fig: 3gppNetSlicingArch}
	\end{center}
\end{figure*}

The 3GPP network slicing framework does not inherently support inter-slice 
handovers. That is, the session continuation support among slices is 
not 
specified. As a result, when a user/UE wishes to change its slice, its ongoing 
session at its current slice is released before it can be re-established over 
an alternate slice. Apart from that, mechanisms for inter-slice handover 
decision are also necessary, among other key features. In this vein, this 
article aims to identify key technical gaps and challenges for inter-slice 
mobility as per the current 3GPP specifications (Release 16). Hereinafter, we 
use the term 
\textit{inter-slice switching} to refer to the process of UE switching (or 
changing) slices without session continuation. The terms \textit{inter-slice 
handover} 
and \textit{inter-slice mobility management} are used interchangeably to refer 
to the 
process when a UE moves among slices with seamless continuation of its 
ongoing session. 

In the following, we first discuss the 3GPP standard principles and relevant 
mechanisms that constitute inter-slice switching in SBA. Several possible 
forms (referred to as \textit{cases}) of inter-slice mobility are discussed. A 
detailed 
explanation of one example case is provided to describe the underlying 
procedures. Finally, some challenges, and the corresponding research 
directions are identified, which can be pursued for developing comprehensive and 
efficient 
inter-slice handover solutions within the standard network slicing framework of 
the 3GPP.

\section*{Inter-Slice Switching in 3GPP SBA - Standard Principles}

A network slice, according to 3GPP, is a logical network with specific 
network characteristics and capabilities. It is essentially a set of 
virtual/logical network functions (NFs) that run on top of network resources 
such as compute, storage, and networking. These NFs can be overarching (e.g., 
NSSF), slice-specific (e.g., SMF/UPF), or shared among slices. Some NFs, 
however, can be deployed flexibly either as slice-specific or shared, depending 
on deployment needs. The AMF is a prominent such example  as shown in Fig. 
\ref{fig: 3gppNetSlicingArch}. In addition, instances of some other NFs, such 
as UDM/UDR, 
can possibly exist simultaneously as overarching, slice-specific, as well as 
shared.



In order to communicate over a slice, it is required that a UE first registers 
itself 
with the slice. 
For this purpose, the UE carries out a Registration procedure with the AMF. 
The 
AMF, in addition to Registration Management, is also responsible for 
access control and mobility management for UE. 
After successfully registering with the slice, the UE can 
establish a session with a Data Network (DN) through this slice. The 
traffic 
exchange between the UE and DN is in the form of PDUs (Protocol Data Units), 
and the communication session among them is termed as a PDU session. A PDU 
session can be of type IP, Ethernet, or Unstructured, to support requirements 
of different service types (or use cases) \cite{23501}. The UE sends 
a request  to Session Management Function (SMF)  for PDU session 
establishment (as 
well as PDU session release when required). 
Apart from UE's session management, SMF also configures and controls the User 
Plane Functions (UPFs). A UPF is the data plane entity at the core 
network where the actual traffic routing and forwarding takes place. The role 
of other NFs in inter-slice switching, as shown in Fig. \ref{fig: 
3gppNetSlicingArch}, is described later in the following sections.

\subsection*{Principles for Inter-Slice Switching}

A network slice in SBA is commonly identified through an identifier namely 
S-NSSAI (Single Network Slice Selection Assistance Information).  The 3GPP 
standard procedures in a sliced network usually deal with a 
set of 
S-NSSAIs, which form an \textit{NSSAI}. Every PLMN (Public Land Mobile Network) 
domain supports a specific set of S-NSSAIs for UEs known as 
\textit{Configured NSSAI}. A UE can 
have subscriptions to multiple S-NSSAIs in a network. An S-NSSAI with which a 
UE has an active subscription is referred to as a \textit{Subscribed 
S-NSSAI}. A UE, however, can only avail services of a slice (e.g., 
establishing a PDU session to a DN over it), if the network allows 
connectivity 
over the slice. A set of slices to which 
the UE is allowed to connect to at any given time is termed as 
\textit{Allowed NSSAI}. A UE can access up to eight slices at a time. 

\begin{figure*}[!hbp]
	\begin{center}
		\includegraphics[scale=0.35]{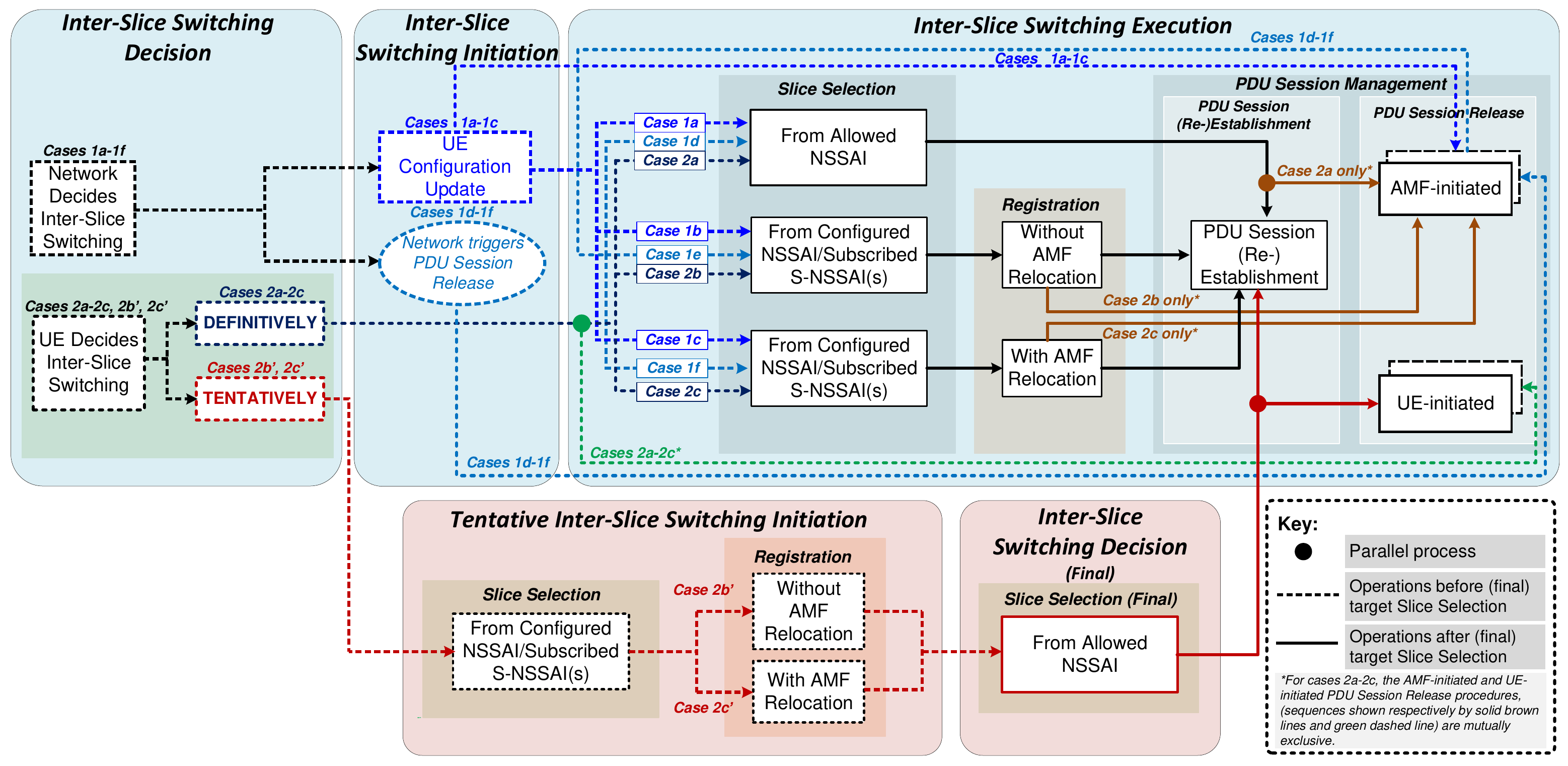}
		\caption{Inter-slice switching cases - sequence of operations}
		\label{fig: BlockDiagramsCombined}
	\end{center}
\end{figure*}

In principle, a UE can only switch to a slice if it is present in its 
\textit{Allowed NSSAI}. If a UE 
wishes to access a slice to which it is subscribed to, but is currently not 
present in the \textit{Allowed NSSAI}, it can request the network to include 
the slice 
in the \textit{Allowed NSSAI} by sending a \textit{Requested NSSAI} through a
Registration procedure (discussed later). A \textit{Requested NSSAI} refers 
to 
the set of slices (S-NSSAIs) requested by the UE to be included in the 
\textit{Allowed NSSAI}. If a UE does not explicitly request the network 
for a 
particular slice (S-NSSAI), the network serves the UE via at least one 
\textit{default S-NSSAI}
(slice), which is chosen from the \textit{Subscribed S-NSSAIs} of the UE.

The inter-slice switching 
may or may not require a modification in the current \textit{Allowed NSSAI}. 
The modification of the \textit{Allowed NSSAI} can be done either by the UE or 
the 
network slice itself, by carrying out certain 
procedures. 
%
The modification of the \textit{Allowed NSSAI} is followed by the 
PDU Session 
Management process, which includes the release of PDU session from the 
current 
slice
and its (re-)establishment with the desired target 
slice. 
Specifically, the 
procedures for the modification of \textit{Allowed NSSAI} involve mechanisms such 
as the UE Configuration Update and Registration, while session 
management involves PDU Session Release and PDU Session (Re-)Establishment 
procedures.

\textbf{UE Configuration Update:} This procedure is normally used by the 
network to 
update certain configurations at the UE side, for instance, Access and Mobility 
Management related parameters. It can also be used to modify the 
\textit{Allowed NSSAI} of the UE. In the context of inter-slice switching, 
the network/AMF can enforce the removal of a slice from \textit{Allowed NSSAI} 
with which the UE has an active session. This will force the 
UE  
to connect to an alternate slice.

\textbf{Registration (with or without AMF Relocation):} Registration is 
normally required when a UE wishes to access network services or moves out 
of a registration area. It can also be used by a UE to request 
modification of the \textit{Allowed NSSAI}. In the context of inter-slice 
switching, the 
UE can carry out Registration in order to acquire the desired slice(s) 
(S-NSSAI(s)) in the \textit{Allowed NSSAI}. During the Registration process, 
the 
AMF Relocation 
may also take place (i.e., a new AMF may be chosen) if the current AMF is 
unable 
to serve all slices in the new \textit{Allowed NSSAI}.

\textbf{UE-/Network-Initiated PDU Session Release:} Through this 
procedure, the network or the UE can initiate the release of an ongoing PDU 
session. In the context of inter-slice switching, 
the network may initiate the 
PDU Session Release to indicate the unavailability of a slice.  
This procedure at the network slice 
can be initiated by the AMF, SMF, or PCF. 
With the PDU Session Release procedure, all 
configurations (e.g., QoS configurations) as well as resources
associated with the PDU Session are released. Such resources include, the 
allocated IP address, any UPF resources, and RAN resources.

\textbf{PDU Session Establishment:} In the context of inter-slice 
switching, the UE carries out the PDU Session Establishment to (re-)establish 
its (ongoing) session over an alternate slice. The UE can decide to 
initiate this procedure itself if it wishes to switch to 
another slice. The UE may also carry out this procedure if it is forced by 
the network to switch slices (i.e., through the aforementioned UE Configuration 
Update or the PDU Session Release procedures).


\section*{Forms of Inter-Slice Switching in 3GPP SBA}
\label{sec: forms}

Depending on the availability of the 
candidate S-NSSAI and the PDU session status, both the UE-initiated inter-slice 
switching and the network-triggered inter-slice 
switching can occur in
several forms. These forms accordingly define the order of sequence of their 
respective protocol operations. We refer to these forms 
as different \textit{cases} of inter-slice switching. These are briefly described below. The sequence of 
the involved procedures in each case is shown in Fig. \ref{fig: 
BlockDiagramsCombined}.


\subsection*{Network-Triggered Inter-Slice Switching}

Cases 1a to 1f represent network-triggered inter-slice switching. Cases 
1a, 1b, and 1c  are triggered through UE Configuration Update, while Cases 
1d, 1e, and 1f are triggered by enforcing the PDU Session Release. An 
AMF-initiated PDU Session Release procedure is also carried out in Cases 1a, 
1b, and 1c, as a result of UE Configuration Update (Fig. \ref{fig: 
BlockDiagramsCombined}). This is because the AMF determines that the slice with 
an active PDU session with UE is now unavailable in its \textit{Allowed NSSAI}. 

In Cases 1a and 1d, the UE is able to choose a suitable alternate slice from 
the already available \textit{Allowed NSSAI}. The UE can then (re-)establish 
its session over this slice. In other cases, however, the UE does not have a 
suitable alternate slice in 
the \textit{Allowed NSSAI} to connect to. Therefore, it chooses the alternate 
slice (S-NSSAI) from the \textit{Configured NSSAI/Subscribed 
S-NSSAI(s)} and performs Registration to obtain its desired slice 
(S-NSSAI) in 
\textit{Allowed NSSAI}. The Registration process in Cases 1b and 1e does not 
require the AMF 
Relocation/(Re-)selection. However, for Cases 1c and 1f, the Registration 
process also
involves the AMF Relocation. 

\subsection*{UE-Initiated Inter-Slice Switching}
%
In contrast to the network-triggered inter-slice switching,
the UE can possibly choose to initiate the inter-slice switching
\textit{tentatively} or \textit{definitively}. Cases 2a to 2c shown in Fig. 
\ref{fig: 
	BlockDiagramsCombined}
are \textit{definitive} cases, while  Cases 2b’ and 2c’ are 
\textit{tentative}
cases. In \textit{definitive} cases, the UE decides to switch slices
definitely (i.e., it decides to leave the current slice regardless
of whether a suitable alternate slice is available in \textit{Allowed
NSSAI}, e.g., due to very high costs or zero throughput etc.). Accordingly, the 
PDU Session Release procedure is also triggered either right away by the UE 
(sequence represented through dashed green line in Fig. \ref{fig: 
BlockDiagramsCombined}), or by the network during the PDU Session 
(Re-)Establishment in Case 2a, or Registration in Cases 2b and 2c (sequences 
represented through solid brown lines in Fig. \ref{fig: 
BlockDiagramsCombined}). In \textit{tentative} cases, on the other hand, the 
user does 
not experience any unacceptable
issues with the current slice. It simply attempts to obtain a (set of)
possible alternate slice(s) (S-NSSAI(s)) in \textit{Allowed NSSAI} through
Registration (e.g., for same service guarantees at lower costs).
It makes the final slice selection and decides to switch slices only after the
successful completion of the Registration process.

It is worth mentioning that in the \textit{tentative} cases, the modification to
\textit{Allowed 
NSSAI} during Registration does not remove the currently active slice (S-NSSAI) 
from the \textit{Allowed NSSAI}. So, whether the Registration process completes 
successfully or not, the PDU session of UE over the current slice remains 
intact until the UE makes the final decision to switch slices.

%
In the \textit{definitive} Case 2a, the UE decides to switch to an 
alternate 
slice that 
is already 
present in \textit{Allowed NSSAI}. In Cases 2b and 2c, the UE first performs 
Registration to obtain its target slice 
(S-NSSAI) in \textit{Allowed NSSAI}. For Case 2b, the Registration 
does not require the AMF Relocation. For Case 2c, the Registration does require
AMF Relocation. During Registration, as soon as the AMF 
learns that the modification to \textit{Allowed NSSAI} has led to the 
unavailability of a currently active slice, it 
initiates the PDU Session Release procedure over this slice as well. 
Notably, 
such initiation 
of PDU Session Release during Registration does not occur in 
network-triggered inter-slice switching cases. This is because in each of those 
cases the PDU 
Session Release is already executed before Registration either explicitly 
(i.e., 
for Cases 1d, 1e, and 1f) or on successful completion of the UE Configuration 
Update procedure (i.e., for Cases 1a, 1b, and 1c).

The \textit{tentative} cases 2b' and 2c' also follow the same sequence of 
procedures as 
2b and 2c, however, unlike Cases 2b and 2c, the Registration in Cases 
2b' and 2c' does not remove the currently active slice (S-NSSAI) from the 
\textit{Allowed NSSAI}. This allows the UE to make the final decision to 
switch slices after the Registration process completes successfully. 

\begin{figure*}[!htbp]
	\begin{center}
		\includegraphics[scale=0.52]{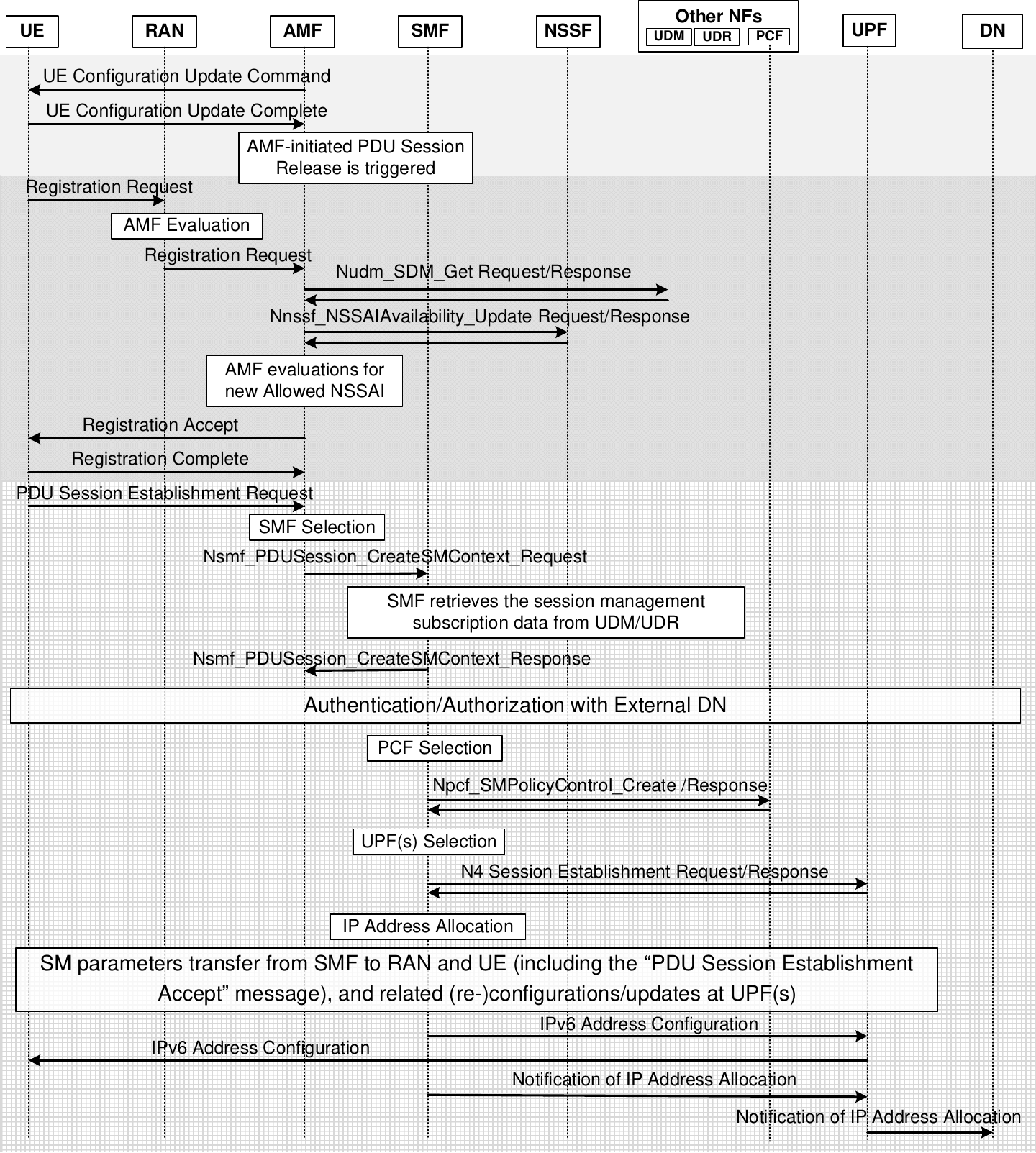}
		\caption{Signaling sequence for inter-slice switching operation (Case 
		1b)}
		\label{fig: SignalingSequence}
	\end{center}
\end{figure*}

\subsection*{An Example Inter-Slice Switching Case}
\label{subsec: exampleISHOcase}

We now summarize the  workflow of an example inter-slice switching case to show 
the role and interaction among different SBA NFs during the inter-slice 
switching process. Case 1b is chosen for this purpose as its operation
encompasses most major procedures common in some other
cases as well. The signaling sequence of Case 1b is shown
in Fig. \ref{fig: SignalingSequence}. The constituent procedures of Case 1b 
including the UE Configuration Update, PDU Session Release, Registration and 
PDU Session Establishment are described as specified in the 3GPP standard TS 23.502 \cite{23502}.  

%
%

In Case 1b, the AMF triggers inter-slice switching by removing the currently 
active slice (S-NSSAI) of UE from its \textit{Allowed NSSAI}. The UE’s session 
is 
released, and 
it stops receiving traffic from the current slice. The AMF communicates the 
modified \textit{Allowed NSSAI} to the UE via UE Configuration Update Command 
message. 
The UE, however, chooses its alternate slice from \textit{Configured 
NSSAI/Subscribed 
S-NSSAI(s)}. The UE then sends a Registration Request message containing its 
\textit{Requested 
NSSAI} to AMF. The AMF verifies the \textit{Requested NSSAI} through UE’s 
subscription 
information, which it retrieves from UDM/UDR. The NSSF can also assist the AMF 
for 
\textit{Requested NSSAI} verification, and provisioning of new \textit{Allowed 
NSSAI}. After 
verification, the AMF sends the new \textit{Allowed NSSAI} with UE’s desired 
slices 
(S-NSSAI(s)) in the Registration Accept message. 

The UE is now ready to start the PDU Session \linebreak(Re-)Establishment with 
its 
desired slice (S-NSSAI). It sends the PDU Session Establishment Request to SMF 
via AMF. The AMF may first select a suitable SMF, especially if the operator 
deploys multiple SMFs (e.g., for load balancing). To process 
the UE’s request, the SMF first retrieves the Session Management (SM) 
subscription data 
from UDM. After verifying the UE’s SM subscription, the SMF performs a number 
of functions before accepting the UE’s PDU Session Establishment request. These 
include:

\begin{itemize}
	\item Initiation of UE authentication/authorization with external DN.
	\item PCF selection and retrieval of the policy information (e.g., charging 
	and QoS information) from PCF. This information is enforced by the SMF 
	during the PDU Session Management.
	\item UPF(s) selection, which will handle UE’s traffic at data plane. N4 
	sessions establishment with UPF(s) also takes place which allow SMF-UPF 
	interaction. During the N4 sessions establishment, the SMF provides UPF(s) 
	with packet detection, enforcement, and reporting rules for handling the 
	UE’s traffic at the data plane.
	\item IP address allocation, which (assuming IPv6 addressing) is 
	advertised to UE later on when the PDU Session Establishment completes 
	successfully.
	\item Communicating the SM parameters to UE and RAN.
\end{itemize}



The successful configuration of the SM parameters at UE and RAN also marks the
completion of the PDU session establishment process. The SMF eventually 
provides the IPv6 Address Configuration information (i.e., an IPv6 Prefix) and 
sends it to the UE via the UPF(s). The uplink/downlink packet delivery from/to 
the UE subsequently starts over the new slice.

\section*{Towards Seamless Inter-Slice Mobility -- Key Challenges and 
Research Directions}
\label{sec: futureworks}

	The capability to ensure seamless inter-slice mobility is 
	an essential 
requirement in a sliced mobile network. 
It is thus
imperative that the inter-slice switching mechanisms are enhanced with 
seamless inter-slice handover support mechanisms.  
In this vein, some key technical gaps and challenges, as well as
the corresponding research directions are discussed as follows.

\textit{\textbf{Session Continuity:}} 	Smooth continuation of an ongoing 
session is a primary requirement to achieve an inter-slice handover.
The session continuation for IPv6 
sessions is 
considered in \cite{stpaper}, where session continuation among slices is 
achieved through the standard Mobile IPv6, and the GPRS Tunnelling Protocol 
(GTP) 
of 
3GPP. These solutions are shown to impose a trade-off between low latency, and 
higher (signaling and resource utilization) costs. 
The potential alternate approach is network-based session 
continuation mechanisms (e.g., based on Proxy Mobile IPv6 principles specified 
in the Internet Engineering Task Force RFC 5213), which can balance such 
trade-offs \cite{lee2013}. 

\textit{\textbf{Timely Slice Selection and Inter-Slice Handover Triggering:}} 
For an efficient inter-slice handover operation, it is critical that a suitable 
target slice is decided in a timely manner, and handover is triggered to the 
target slice at a precise instant (i.e., neither too early, nor too late). 
Both these are complex challenges considering the dynamics of a network-sliced 
environment. This complexity becomes more evident when the candidate slices are 
orchestrated based on dynamically shared resources. A powerful approach to 
address these challenges is to use data analytics, which paves the way to apply 
machine learning techniques. Several machine learning techniques have already 
been proposed which can effectively predict parameters such as network delay, 
loss rate, jitter, throughput etc. \cite{MLSDNSurvey}. Based on these predicted 
values, a suitable target slice among several candidate slices can be selected 
and time-to-handover towards it can also be determined.

\textit{\textbf{New Protocol Entities for Inter-Slice Handover Management:}} 
The inter-slice handover being an enhanced capability requires new 
protocol entities in the 3GPP network slicing framework (e.g., for target slice 
selection, and others discussed next). For network-triggered inter-slice 
handovers, these functionalities can be executed at an inter-slice handover 
manager which can be defined as a dedicated overarching NF at SBA. This NF can 
utilize 
functionalities of other core NFs, for instance, it can leverage a cross-slice, 
overarching NWDAF -- the standard NF for data analytics -- for any decision 
making capabilities, or NSSF for target 
slice 
selection. A potential alternate is to deploy the inter-slice handover 
manager as a dedicated management function at the management plane. The 
emerging Artificial 
Intelligence-based management systems such as Zero-Touch Network and Service 
Management (ZSM) offer several features for effective management of inter-slice 
handover 
related operations, even in highly dynamic and complex network slicing 
environments 
\cite{BenzaidZSM2020}. The inter-slice handover manager at ZSM can utilize 
the standard ZSM 
services, for instance, its end-to-end analytics and intelligence services for 
inter-slice handover decision making etc. 

For UE-initiated inter-slice handovers, however, enhancing user devices or UEs 
with new protocol entities is not straightforward. This is mainly due to 
challenges involved in the required modifications in the UE’s protocol stack. 
Moreover, these functionalities may be resource-intensive causing significant 
overheads on the limited UE resources (e.g., battery power). Fortunately, 
middleware solutions exist which can act as handover managers on behalf of UE 
\cite{Nasser2007}. These managers (e.g., hosted at a nearby trusted fog server) 
can  run 
computationally intensive tasks (e.g., machine learning algorithms) on UE’s 
behalf during the inter-slice handover process.

\textit{\textbf{Inter-Slice Handover Information Gathering and Exchange:}} 
The effectiveness of data analytics and machine learning at the inter-slice 
handover manager entities requires timely and up-to-date information on the 
prevailing 
conditions of the target/candidate slices. At SBA, the slice-specific 
NWDAF can be seen as a central entity for inter-slice handover related 
information gathering and exchange. It can receive various events information 
from other core network NFs such as AMF, SMF, and PCF. Information/Data 
retrieval from 
other NFs such as UDM/UDR, NSSF, and NRF, is also possible. Although the 
inter-slice handover manager NF at SBA can leverage information from NWDAF of 
each slice, the standard NWDAF interactions are confined only to core network 
NFs of a 
slice. For operations such as inter-slice handover decision, RAN information 
from target slices as well as from UE is also desired. For this purpose, 
enhancements to existing slice information gathering and exchange mechanisms 
are necessary. In fact, some works (e.g., \cite{E2EDataAnalytics}) have already 
proposed solutions in this direction which extend the existing network data 
analytics framework and its interactions beyond the core network NFs, 
encompassing, for 
instance, the RAN and management plane as well.

For ZSM-based inter-slice handover manager, the standard ZSM data collection 
services offer 
additional  advantages. In addition to supporting data collection from RAN and 
core network domains, these services can collect infrastructure-level 
information as 
well (e.g., about resource-consumptions of individual NFs from the underlying 
Network Functions Virtualization (NFV) Orchestrator). 

The UE based inter-slice handover manager would mostly rely on user-perceived 
parameters such 
as Quality-of-Experience, throughput, delay, loss rate etc. However, 
information from candidate 
slices is also necessary. Slice advertisement mechanisms (e.g., as proposed in 
\cite{XAn2017}) can be implemented to provide such information to the UE-based 
inter-slice handover manager.


\textit{\textbf{Inter-Slice Handover Preparation:}} The ability to predict the 
inter-slice handover beforehand can foster mechanisms for inter-slice 
handover 
preparation. Slice advertisements containing up-to-date slice information is an 
example of inter-slice handover preparation mechanism. These mechanisms, in 
turn, can support the proactive initiation of inter-slice handovers (e.g., 
proactive establishment of an inter-slice tunnel \cite{stpaper}) as opposed to 
triggering the 
handover reactively as the conditions have already deteriorated.   


\textit{\textbf{Inter-Slice Handover Scheduling:}} The inter-slice handover 
process may not be a desirable operation for UEs belonging to URLLC and 
V2X service types. In this regard, the UEs or slice owners may 
\textit{schedule} the inter-slice handover process at specific intervals only. 
For instance, when a vehicle (a V2X UE) is stationary or moving in a 
non-congested area, or when a robotic device at a remote factory (a URLLC UE) 
is performing a non-critical task. Again, the prediction mechanisms can provide 
necessary intelligence to schedule the execution of inter-slice handover at a 
suitable interval.


\textit{\textbf{Managing Inter-Slice Handovers with Horizontal and Vertical 
Handovers:}} A critical mobility management scenario in a network-sliced 
environment occurs when an inter-slice handover is triggered as a result of a 
horizontal or a vertical handover (as depicted in Fig. \ref{fig: Scenario}). 
Both these scenarios are prone to high latencies as they require simultaneous 
management of a UE's mobility to a new subnet or access technology and to 
a new slice. Hence, in addition to the standalone inter-slice handover 
solutions, integrated solutions would be needed which can collectively handle 
horizontal/vertical  handovers alongside inter-slice handovers within a unified 
mobility management framework.


\textit{\textbf{Security of the Inter-Slice Handovers:}} The security threat 
landscape of network slicing is extremely broad and constantly evolving. This 
is due to the embodiment of various technologies such as Software-Defined 
Networking, NFV, Internet-of-Things, Machine 
Learning etc. \cite{Olimid2020} – each bringing its own set of vulnerabilities 
into network 
slicing. Securing inter-slice handover is thus a critical requirement. A number 
of security attacks can be launched by exploiting the handover signaling 
messages between the UE and slices. The clear text transmission of 
\textit{Allowed/Requested NSSAI} in these messages makes the inter-slice 
handover 
process particularly vulnerable to several threats. The possible threats 
include Denial-of-Service, Session Hijacking, Malicious Mobile Node Flooding, 
Man-in-the-middle, and redirection attacks. For example, an authorized but 
malicious node 
masquerading as an AMF, can transmit fake \textit{Allowed NSSAI} in a false UE 
Configuration Update message to a set of UEs (e.g., IoT devices). This can 
prompt these devices to simultaneously send PDU Session (Re-)Establishment or 
Registration Request messages to a target slice. Overloading a slice with such 
requests can cause Denial-of-Service for legitimate users. Potential approaches 
to 
mitigate such threats include privacy protection mechanisms for S-NSSAIs, for 
example, 
the use of \textit{encrypted S-NSSAI}, or replacing the actual S-NSSAIs in 
signaling 
messages 
with \textit{Temporary S-NSSAIs} as studied in the 3GPP technical report 33.813.


\section*{Conclusions}
	Representing a paradigm shift in network engineering, 
	network slicing requires 
new protocols. In particular, mobility management in a 
network-sliced environment requires new and efficient solutions. This article 
investigates the problem of inter-slice mobility from the 3GPP  standards’ 
perspective. It provides a thorough overview of the 3GPP standard  
principles pertinent to the UE's movement between different slices. Based on 
these 
principles, the article highlights some prospective 
research directions, and in particular, the potential of “data analytics” and 
“machine learning” techniques for supporting seamless 
inter-slice mobility, 
consistent with the current 3GPP network slicing framework. 
\ifCLASSOPTIONcaptionsoff
  \newpage
\fi

\begin{IEEEbiographynophoto}{Muhammad Mohtasim Sajjad}
[S'17] (m.sajjad@ieee.org) is a PhD candidate with the  School of Electrical 
Engineering and Robotics, 
Queensland University of Technology, Brisbane, 
Australia. His research interests include network slicing, 
network virtualization and mobile communications.  
\end{IEEEbiographynophoto}

\begin{IEEEbiographynophoto}{Carlos J. Bernardos}
(cjbc@it.uc3m.es) is an Associate Professor at the University Carlos III of 
Madrid, Spain.  
His current research interests include network virtualization 
and 5G.
\end{IEEEbiographynophoto}

\begin{IEEEbiographynophoto}{Dhammika Jayalath}
[M'99, SM'11] (dhammika.jayalath@qut.edu.au) is a Senior Lecturer with the 
School of 
Electrical Engineering and Robotics, 
Queensland University of Technology, Brisbane, 
Australia. His current research interests include 5G systems, cooperative 
communications, communications theory. 
\end{IEEEbiographynophoto}

\begin{IEEEbiographynophoto}{Yu-Chu Tian}
[M'00, SM'19] (y.tian@qut.edu.au) is a Full Professor with the School of 
Computer Science, Queensland University of 
Technology, Brisbane, Australia. His current research 
interests include big data computing, cloud computing, computer networks, optimization and machine learning, networked control systems, and cyber-physical system security.  
\end{IEEEbiographynophoto}



\end{document}